\documentclass[11pt]{article}

\usepackage{amssymb,amsmath,amsthm}
\usepackage{bbm}
\usepackage[noend]{algpseudocode}
\usepackage{graphicx}
\usepackage{verbatim}%for comment blocks
\usepackage{url,xspace,hyperref}
\usepackage{cite}
\usepackage{caption}
\usepackage{subcaption}
\usepackage{multirow}
\usepackage{multicol}
\usepackage{latexsym}
\usepackage{amsmath,amssymb,enumerate}
\usepackage{algorithm,algorithmicx}
\usepackage{float}
\usepackage{xcolor}
\usepackage{mathrsfs}
\usepackage{cleveref}
\usepackage{booktabs}
\usepackage{breqn}

\usepackage{fullpage}
\usepackage{geometry}
\usepackage{setspace}
\usepackage{tcolorbox}
\usepackage{xspace}

\tcbuselibrary{skins,breakable}
\tcbset{enhanced jigsaw}

\newtheorem{theorem}{Theorem}[section]
\newtheorem*{theorem*}{Theorem}
\newtheorem{defn}{Definition}[section]

\newcounter{note}[section]

\def\sse{\subseteq}
\newcommand{\dom}{\mathsf{dom}}

\def\asc{\ensuremath{{\sf ASC}}\xspace}

\newcommand{\E}{\mathbb{E}} 
\def \Z{\mathbb Z}

\def \R{\mathbb R}
\DeclareMathOperator*{\argmax}{arg\,max}

\newcommand{\ignore}[1]{}

\allowdisplaybreaks

\title{Lower Bound on the Greedy Approximation Ratio \\for Adaptive Submodular Cover} 

\author{Blake Harris\thanks{Department of Industrial and Operations Engineering, University of Michigan, Ann Arbor, USA. Research supported in part by NSF grant CCF-2006778.}  \and Viswanath Nagarajan$^*$}
\begin{document}

\maketitle

\thispagestyle{empty}
\begin{abstract} 
We show that the greedy algorithm for adaptive-submodular cover has  approximation ratio at least $1.3\times (1+\ln Q)$.  Moreover, the instance demonstrating this gap has $Q=1$. So, it invalidates a prior result in the paper ``Adaptive Submodularity: A New Approach to Active Learning and Stochastic Optimization'' by Golovin-Krause, that  claimed a $(1+\ln Q)^2$-approximation ratio for the same algorithm.
\end{abstract}

\section{Introduction} 
Adaptive-submodularity is a widely used framework in stochastic optimization and machine learning~\cite{GK11,GK17,EKM21,ACN}. Here, an  algorithm makes sequential decisions while (partially) observing uncertainty. We study a basic problem in this context:  covering an adaptive-submodular function at the minimum expected cost. We show that the natural greedy algorithm for this problem has approximation ratio at least $1.3\times (1+\ln Q)$, where $Q$ is the maximal function value. This is in contrast to   special cases such as  deterministic submodular cover or (independent) stochastic submodular cover, where the greedy algorithm achieves a tight  $(1+\ln Q)$ approximation ratio. 

\subsection{Problem Definition}\label{sec:Definition}

The following definitions and problem description are borrowed from  \cite{ACN}.    

\paragraph{Random items.} Let  $E$ be a finite set of $n$ {\em items}. Each item $e \in E$ corresponds to a random variable $\Phi_e \in \Omega$, where $\Omega$ is the {\em outcome space} (for a single item). We use $\Phi= \langle \Phi_e : e\in E\rangle$ to denote the vector of all random variables (r.v.s). The r.v.s may be arbitrarily correlated across items.   
We use upper-case letters to represent r.v.s and the corresponding lower-case letters to represent realizations of the r.v.s. Thus, for any item $e$, $\phi_e\in \Omega$ is the realization of $\Phi_e$; and $\phi=\langle \phi_e : e\in E\rangle$ denotes the realization  of $\Phi$.
 Equivalently,  we can represent the realization $\phi$  as a subset $\{(e,\phi_e): e\in E\}\sse E\times \Omega$ of item-outcome pairs.

 A {\em partial realization} $\psi\sse E\times \Omega$ refers to the realizations of any {\em subset} of items;  $\dom(\psi)\sse E$ denotes the items whose realizations are represented in $\psi$, and $\psi_e$ denotes the realization of any item $e\in \dom(\psi)$. Note that a partial realization contains at most one pair of the form $(e,*)$ for any item $e\in E$. The (full) realization $\phi$ corresponds to a partial realization with $\dom(\phi)=E$. For two partial realizations $\psi,\psi'\sse E\times \Omega$, we say that $\psi$ is a \textit{subrealization} of $\psi{'}$ 
 (denoted $\psi\preccurlyeq\psi{'}$)  
 if $\psi\subseteq\psi'$; in other words, $\dom(\psi)\sse \dom(\psi')$ and $\psi_e=\psi'_e$ for all $e\in \dom(\psi)$.\footnote{We use the notation $\psi\preccurlyeq\psi{'}$ instead of $\psi\sse\psi{'}$ in order to be consistent with prior works.} 
 Two partial realizations $\psi,\psi'\sse E\times \Omega$ are said to be {\em disjoint} if there is no full realization $\phi$ with $\psi \preccurlyeq\phi$ and $\psi' \preccurlyeq\phi$; in other words, there is some item $e\in \dom(\psi)\cap \dom(\psi')$ such that the realization of $\Phi_e$ is different under $\psi$ and $\psi'$.
 
 We assume that there is a prior probability distribution $p(\phi)=\Pr[\Phi=\phi]$ over realizations $\phi$. Moreover, for any partial realization $\psi$, we assume that we can compute the posterior distribution $p(\phi|\psi) = \Pr(\Phi=\phi|\psi\preccurlyeq\Phi).$

\paragraph{Utility function.} 
 In addition to the random items (described above), there is a {\em utility function} $f : 2^{E\times\Omega} \to \R_{\geq 0}$ that assigns a value to any partial realization.  We will assume that this function is monotone, i.e., having more realizations can not reduce the value. Formally, 
 \begin{defn}[Monotonicity]\label{Definition:mon}
A function $f : 2^{E\times\Omega} \to \R_{\geq 0}$ is {\bf monotone} if 
$$f(\psi)\le f(\psi')\quad \mbox{ for all  partial realizations }\psi\preccurlyeq\psi{'}.$$
\end{defn}

We also assume that the function $f$ can always achieve its maximal value, i.e.,

\begin{defn}[Coverable]\label{Definition:coverable}
Let $Q$ be the maximal value of function $f$. Then, function $f$ is said to be {\bf coverable} if this value $Q$ can be achieved under every (full) realization, i.e., 
\[f(\phi)=Q \mbox{ for all realizations $\phi$ of  }\Phi.\]
\end{defn}

Furthermore, we will assume that the function $f$ along with the probability distribution $p(\cdot)$ satisfies a submodularity-like property. Before formalizing this, we need the following definition. 
\begin{defn}[Marginal benefit]  The {\bf conditional expected marginal benefit}  of an item $e\in E$ conditioned on observing the partial realization $\psi$ is:
\[\Delta(e|\psi):=\E\left[f(\psi \cup (e, \Phi_e)) -f( \psi) \,|\, \psi\preccurlyeq\Phi\right] \,\,=\,\, \sum_{\omega\in \Omega}\Pr[\Phi_e=\omega|\psi\preccurlyeq\Phi]\cdot \left( f(\psi\cup (e,\omega) )-f(\psi)\right).\]
\end{defn}

We will assume that function $f$ and distribution $p(\cdot)$ jointly satisfy the adaptive-submodularity property, defined as follows. 
\begin{defn}[Adaptive submodularity]\label{Definition:adsub}
        A function $f : 2^{E\times\Omega} \to \R_{\geq 0}$ is {\bf adaptive submodular} w.r.t. distribution $p(\phi)$ if for all partial realizations $\psi\preccurlyeq\psi^{'}$, and all items $e \in E\setminus\dom(\psi^{'})$, we have
\[\Delta(e|\psi)\geq \Delta(e|\psi^{'}).\]
\end{defn}

In other words, this property ensures that the marginal benefit of an  item never increases as we condition on more realizations. 
Given any function $f$ satisfying Definitions~\ref{Definition:mon}, \ref{Definition:coverable} and \ref{Definition:adsub},  we can pre-process $f$ by subtracting $f(\emptyset)$, to get an equivalent function (that maintains these properties), and has a smaller $Q$ value. So, we may assume that $f(\emptyset)=0$.

\def\cp{c_{p}}

\paragraph{Min-cost adaptive-submodular cover (\asc).} 

In this problem, each item $e\in E$ has a positive cost $c_e$. The goal is to select items (and observe their realizations)  sequentially until the observed realizations have function value $Q$.  The objective is to minimize the expected cost of selected items. 

Due to the stochastic nature of the problem, the solution concept here is much more complex than in the deterministic setting (where we just select a static subset). In particular, a solution  corresponds to a ``policy'' that maps observed realizations to the next selection decision. 
The observed realization at any point corresponds to a partial realization (namely, the realizations of the items selected so far). 
 Formally, a {\em policy} is a mapping $\pi:2^{E\times\Omega}\to E$, which specifies the next item $\pi(\psi)$ to select when the observed realizations are $\psi$.\footnote{Policies  and utility functions  are not necessarily defined over all subsets $2^{E\times \Omega}$, but only over partial realizations;  recall that a  partial realization is of the form $\{(e,\phi_e): e\in S\}$ where $\phi$ is some full-realization and $S\sse E$.}  
The policy $\pi$ terminates at the first point when $f(\psi)=Q$, where $\psi\sse E\times \Omega$ denotes the observed realizations so far. For any policy $\pi$ and full realization $\phi$, let $C(\pi , \phi )$ denote the total cost of items selected by policy $\pi$ under realization $\phi$. Then, the expected cost of   policy $\pi$ is:
$$c_{exp}(\pi) \,\,=\,\, \E_{\Phi}\left[C(\pi, \Phi)\right] \,\,=\,\, \sum_\phi p(\phi)\cdot C(\pi, \phi).$$

At any point in policy $\pi$, we refer to the cumulative cost incurred so far as the {\em time}. If $J_1, J_2,\cdots J_k$ denotes the (random) sequence of items selected by $\pi$ then for each $i\in \{1,2,\cdots k\}$, we view item $J_i$ as being selected during the time interval $[\sum_{h=1}^{i-1} c(J_h) \, ,\, \sum_{h=1}^{i} c(J_h) )$ and the realization of $J_i$ is only observed at time $\sum_{h=1}^{i} c(J_h)$. For any time $t\ge 0$, we use $\Psi(\pi,t)\sse E\times \Omega$ to denote the (random) realizations that have been observed by time $t$ in policy $\pi$. We note that $\Psi(\pi,t)$ only contains the realizations of items that have been {\em completely} selected by time $t$. Note that the policy terminates at the earliest time $t$ where $f(\Psi(\pi,t))=Q$.

\paragraph{Remark:} Our definition of the utility function $f$ is slightly more restrictive than the original definition \cite{GK11,GK17}. In particular, the utility function in \cite{GK11} is of the form $g:2^E\times \Omega^E\rightarrow \R_{\ge 0}$, where the function value $g(\dom(\psi), \Phi)$ for any partial realization $\psi$ is still random and can depend on the outcomes of unobserved items, i.e., those in $E\setminus \dom(\psi)$. We note that the \asc formulation satisfies  ``strong adaptive monotonicity'', ``strong adaptive submodularity'' and the ``self certifying'' condition defined in \cite{GK17}

\def\grd{\pi}
\def\opt{\sigma}
\subsection{Adaptive Greedy Policy}
Algorithm~\ref{alg:THE_algo} describes a natural greedy policy for min-cost adaptive-submodular cover, which has also been studied in prior works \cite{GK17,EKM21,HellersteinKP21,ACN}.

\begin{algorithm}[h!]
\caption{Adaptive Greedy Policy $\grd$.\label{alg:THE_algo}}
\begin{algorithmic}[1]
\State selected items $A \gets \emptyset$, observed realizations $\psi \gets \emptyset$
\While{$f(\psi)<Q$}
\State $e^{*} = \argmax_{e\in E\setminus A}\frac{\Delta(e|\psi)}{c_e}$
\State add $e^*$ to the selected items, i.e., $A \gets A\cup\{e^{*}\}$
\State select $e^*$ and observe $\Phi_{e^*}$
\State update $\psi \gets \psi\cup\{(e^{*}, \Phi_{e^{*}})\}$
\EndWhile
\end{algorithmic}
\end{algorithm}

\subsection{Prior Results on the Greedy Policy}
The above greedy policy  is known to have strong performance guarantees for adaptive submodular cover  and even better performance in  several special cases. We now discuss these results in detail. Here, we assume that the function $f$ is integer valued (to keep notation simple). 

In the {\em deterministic} setting of submodular cover (which is a classic problem in combinatorial optimization), \cite{W82} proved that this greedy algorithm achieves a $1+\ln Q$ approximation. Previously, such an approximation ratio was known in the special case of set cover \cite{Chvatal79}, where $Q$ equals the number of elements to cover. Furthermore, assuming $P\ne NP$, there one cannot achieve any $(1-o(1)) \ln Q$  approximation ratio for set cover~\cite{DinurS14}.

The {\em independent} special case of adaptive-submodular cover (called {\em stochastic submodular cover}) has itself  been studied extensively. Here,   the random variables $\Phi_e$ are independent across items $e\in E$. The greedy policy was shown to achieve an $O(\ln Q)$ approximation ratio  in \cite{INZ12}, and 
recently \cite{HellersteinKP21} proved that the greedy policy has a sharp $(1+ \ln Q )$ approximation guarantee.

Moving beyond the independent setting, \cite{GK11} introduced  adaptive-submodular cover (in a slightly more general form than \asc) and claimed that the greedy policy has a $(1+ \ln Q )$ approximation ratio. This proof had a flaw, which was pointed out by \cite{NS17}. Subsequently, \cite{GK17} posted an updated result proving  a $(1+ \ln Q )^2$ approximation ratio. Then, \cite{EKM21} proved a bound of $(1+ \ln (n Q c_{max}))$ for the greedy policy; note that  this bound depends additionally on the number of items $n$ and their maximum cost $c_{max}$. Recently, \cite{ACN} proved that the greedy policy is a $4\cdot (1+ \ln Q )$ approximation algorithm.

\subsection{Our Result}
In light of the above results, it was natural to expect that the greedy policy for \asc should also have  a $(1+ \ln Q )$ approximation ratio.  This would match the bounds known in the deterministic~\cite{W82} and independent~\cite{HellersteinKP21} special cases.  It would also match the in-approximability threshold known in the (very) special case of set cover~\cite{DinurS14}. However, we show that the greedy policy for \asc cannot achieve a 
$(1+ \ln Q )$ approximation ratio. Specifically,
\begin{theorem}\label{thm:hard-inst}
    There are instances of adaptive submodular cover with integer-valued function $f$ and $Q=1$ where the greedy policy has expected cost  at least $\rho > 1$ times the optimum. 
\end{theorem}
In the next section, we provide a small instance (with $n=4$ items) and prove  that the constant $\rho\ge 1.15$. However, this constant factor can be increased using more complex instances and computer-assisted calculations: we could  obtain $\rho\ge 1.3$.  

Our  result also disproves Theorem~40 in \cite{GK17} that states a $(1+ \ln Q )^2$ approximation ratio for the greedy policy for \asc: this is because our ``hard'' instance  has $Q=1$. So, if the result in \cite{GK17} were correct then the greedy policy should be optimal on our instance. 

\section{Hard Instance for Greedy}
\def\ber{\mathsf{Ber}}
We now prove Theorem~\ref{thm:hard-inst}. 

\begin{figure}[h!] 
    \centering
    \includegraphics[width=6cm]{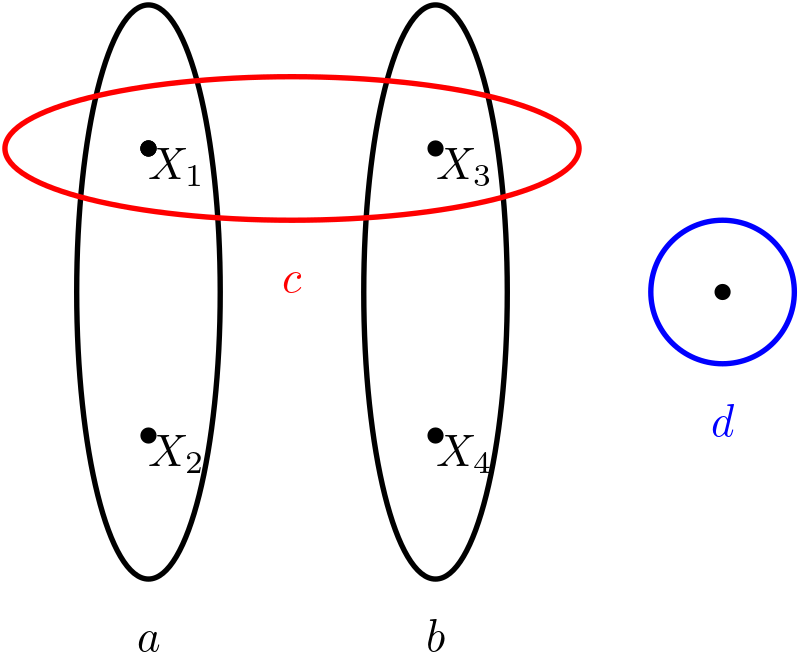}
    \caption{The r.v.s $\{X_i\}_{i=1}^4$ and items $\{a, b , c, d\}$.}
    \label{pic2}
\end{figure}

Our instance consists of $n=4$ items $E=\{a,b,c,d\}$ with outcome space $\Omega=\{0,1\}$. The item realizations are determined by a set of $4$ independent Bernoulli r.v.s $X_1,X_2,X_3,X_4\sim \ber(1-p)$ where the probability $p$ will be set later. We have 
    $$\Phi_a= \begin{cases}
        1 \text{ if } X_1+X_2 \geq 1\\
        0 \text{ otherwise}
    \end{cases}, \quad \Phi_b = \begin{cases}
        1 \text{ if } X_3+X_4 \geq 1\\
        0 \text{ otherwise}
    \end{cases}, \quad \Phi_c= \begin{cases}
        1 \text{ if } X_1+X_3 \geq 1\\
        0 \text{ otherwise}
    \end{cases}.$$ 
    The element $d$ always realizes to $1$, i.e.,  $\Phi_d=1$ with probability $1$. See Figure~\ref{pic2}.

    We set the costs of items $\{a,b,c\}$  to be $1$ each and item $d$ has  cost  $\frac{1}{1-p}$.

The utility function $f:2^{E\times \Omega}\rightarrow \Z_+$ corresponds to observing some ``$1$ outcome''. That is, 
    if we let $E^*=\{(e,1) : e\in E\}$ denote all the $1$-outcome-item pairs then 
    $$f(\psi) = \min\{|\psi \cap E^*|,1\},\qquad \forall \psi\sse E\times \Omega.$$
    Note that $f$ is a coverage function: so it is monotone submodular and has maximal value $Q=1$.

We now show that this instance satisfies all the assumptions for \asc: Definitions~\ref{Definition:mon}, \ref{Definition:coverable} and Definition~\ref{Definition:adsub}.  Clearly, function $f$ is monotone: so Definition~\ref{Definition:mon} holds. Moreover, every full realization $\phi$ has $\phi_d=1$: so $f(\phi)=1=Q$, which means Definition~\ref{Definition:coverable} holds.

\paragraph{Adaptive submodularity.} We now show that  Definition~\ref{Definition:adsub} holds. 
  Suppose $\psi$ and $\psi'$ are   partial realizations  with $\psi\preccurlyeq \psi'$ and item $e\in E\setminus \dom(\psi')$.  
        Now let us examine multiple cases.
        \begin{enumerate} 
            \item $f(\psi')=1$ : in this case, $\Delta(e|\psi')=0$ as $1$ is the maximum value of $f$. So,  $\Delta(e|\psi)\geq\Delta(e|\psi')$.

            \item  $d \in dom(\psi')\cup e$ : if $d\in \dom(\psi')$ then $f(\psi')=1$ and we fall into case~1 above. If $e=d$ then $f(\psi\cup \{(d,\Phi_d)\})=1$, which combined with $f(\psi)\le f(\psi')$ implies $\Delta(e|\psi)\geq\Delta(e|\psi')$.

        \item All remaining cases are summarized in Table~\ref{tab:my_label} or follow using the symmetry between items $a$ and $b$. In each of these cases, we can directly  calculate $\Delta(e|\psi)$ and $ \Delta(e|\psi')$.  E.g., consider the case $\psi=\emptyset$, $\psi'=(a,0)$  and $e=c$: we have   $\Delta(e | \psi) = \Pr[X_1=1 \vee X_3=1] = 1-p^2$ and  $\Delta(e | \psi') = \Pr[X_1=1 \vee X_3=1 | X_1=X_2=0] = 1-p$.    
        \end{enumerate}

        \begin{table} [H]
    \centering
    \begin{tabular}{||c c c c c||}
        \hline
        $\psi$  & $\psi'$  & item $e$ & $\Delta(e | \psi)$ & $\Delta(e | \psi')$\\
        \hline\hline
        $\emptyset$ &$(a,0)$  & $b$ & $1-p^2$& $1-p^2$\\
        \hline
         $\emptyset$ &$(a,0)$ & $c$ & $1-p^2$ & $1-p$\\
        \hline
         $\emptyset$& $(c,0)$&$a$ &$1-p^2$ & $1-p$\\
        \hline
         $\emptyset$&$(a,0) \ (b,0)$ & $c$& $1-p^2$ & $0$\\
        \hline
         $\emptyset$&$(a,0)\ (c,0)$ &$b$ & $1-p^2$ & $1-p$\\
        \hline
         $(a,0)$&$(a,0) \ (b,0)$ & $c$& $1-p$ & $0$\\
        \hline
         $(a,0)$&$(a,0)\ (c,0)$ &$b$ & $1-p^2$ & $1-p$\\
        \hline
    \end{tabular}
    \caption{Remaining cases showing $\Delta(e|\psi)\geq \Delta(e|\psi')$.}
    \label{tab:my_label}
\end{table}

\paragraph{The optimal policy.} It turns out that the optimal policy for this \asc instance selects items in the order $\langle a,  b , d\rangle$ until a 1-outcome is observed. The optimal cost is  
\begin{equation}
    \label{eq:hard-inst-opt}
OPT =1+p^2+\frac{p^4}{1-p}.\end{equation}

 \paragraph{The greedy policy.} We break ties in a specific manner in order to demonstrate a ``bad'' greedy policy below. It is easy to modify the costs slightly so that our instance has a unique greedy policy (which is the one we analyze). For any partial realization $\psi$, we have $\Delta(d|\psi)=1$: so the greedy ratio for item $d$ is always $\frac1{c_d} = 1-p$. For any partial realization $\psi$ that has {\em not} observed at least  one of the r.v.s $\{X_i\}_{i=1}^4$, one of the items $e\in \{a,b,c\}$ has marginal benefit  $\Delta(d|\psi)\ge 1-p$, which means $e$'s greedy ratio is at least $\frac{1-p}{c_e}=1-p$. So, we may assume that   item $d$ will only be chosen when the current partial realization $\psi$ has observed all the underlying Bernoulli r.v.s  $\{X_i\}_{i=1}^4$. 
\begin{enumerate}
    \item Initially, we have partial realization $\psi=\emptyset$ and the greedy ratios are given below: 
\begin{center}
    \begin{tabular}{||c | c | c |c |c||} 
 \hline
item $e$ & $a$ & $b$ & $c$ & $d$  \\ [1ex] 
 \hline
ratio& ${1-p^2}$ & $ {1-p^2}$ & $ {1-p^2}$ & ${1-p}$ \\ [1ex] 
 \hline
\end{tabular}
\end{center}
The greedy policy selects item $c$.

\item Next, the partial realization $\psi=\{(c,0)\}$ with the following ratios. 
\begin{center}
    \begin{tabular}{||c | c| c |c||} 
 \hline
item $e$ & $a$ & $b$ &  $d$  \\ [1ex] 
 \hline
ratio& ${1-p}$ & $ {1-p}$   & ${1-p}$ \\ [1ex] 
 \hline
\end{tabular}
\end{center}
Here, greedy selects item $a$.

\item Next, the partial realization $\psi=\{(c,0), (a,0)\}$ with the following ratios. 
\begin{center}
    \begin{tabular}{||c | c |c||} 
 \hline
item $e$ &  $b$ &  $d$  \\ [1ex] 
 \hline
ratio&   $ {1-p}$   & ${1-p}$ \\ [1ex] 
 \hline
\end{tabular}
\end{center}
Here, greedy selects item $b$.

\item Next, the partial realization $\psi=\{(c,0),(b,0),(c,0)\}$ with only item $d$ remaining. So, greedy selects $d$.
\end{enumerate}
It follows that the greedy policy selects items in the order $\langle c,a,b,d\rangle$ and its expected cost is:
\begin{equation}
    \label{eq:hard-inst-grd}
GRD =1+p^2+p^3+\frac{p^4}{1-p}.
\end{equation}

Using \eqref{eq:hard-inst-opt} and~\eqref{eq:hard-inst-grd}, the greedy approximation ratio is 
$$\frac{{GRD}}{{OPT}}=\frac{1+p^2+p^3+\frac{p^4}{1-p}}{1+p^2+\frac{p^4}{1-p}}$$
 Setting $p=0.7221$, the  approximation ratio is at least $1.15$, which completes the proof of Theorem~\ref{thm:hard-inst}.

\paragraph{Other hard instances.} We can generalize the  above simple instance to a larger class of instances. There are several independent $\ber(1-p)$ r.v.s, denoted ${\cal X}$. There are $n-1$ items, each of which corresponds to a subset   $I\sse {\cal X}$ (the item has outcome $1$ if any of the r.v.s in $I$ is $1$). These $n-1$ items have unit cost. There is also a ``dummy'' item $d$ that always has outcome $1$: this ensures that the instance is coverable (Definition~\ref{Definition:coverable}). The dummy item $d$ has cost $\frac{1}{1-p}$. The function $f$ is the same as before: it is one if a 1-outcome is observed and zero otherwise. Any such instance is adaptive submodular (i.e., Definition~\ref{Definition:adsub} holds).  We could generate several other instances where the greedy policy is not optimal. Using a computer-assisted search, we  also obtained an  instance with greedy approximation ratio $\rho\ge 1.3$.

 \bibliographystyle{alpha}
\bibliography{references}
\end{document}